\documentclass[twocolumn,prb,superscriptaddress,amsmath,amssymb,showpacs]{revtex4-1}

\usepackage{graphicx}
\usepackage{epsfig}
\usepackage{bm}

\begin{document}

\title{Mapping spin-orbit splitting in strained InGaAs epilayers}

\author{B. M. Norman}
\author{C. J. Trowbridge}
\affiliation{Department of Physics, University of Michigan, Ann Arbor, MI 48109}

\author{J. Stephens}
\author{A. C. Gossard}
\author{D. D. Awschalom}
\affiliation{Center for Spintronics and Quantum Computation, University of California, Santa Barbara, CA 93106}
\author{V. Sih}
\affiliation{Department of Physics, University of Michigan, Ann Arbor, MI 48109}
\email{vsih@umich.edu}

\date{\today}

\begin{abstract}
Time- and spatially-resolved Faraday rotation spectroscopy is used to measure the magnitude and direction of the momentum-dependent spin splitting in strained InGaAs epilayers. The epilayers are lattice-matched to the GaAs substrate and designed to reduce inhomogeneous effects related to strain relaxation. Measurements of momentum-dependent spin splitting as a function of electron spin drift velocity along [100], [010], [110] and [1$\overline{1}$0] directions enable separation of isotropic and anisotropic effective magnetic fields that arise from uniaxial and biaxial strain along $\langle$110$\rangle$. We relate our findings to previous measurements and theoretical predictions of spin splitting for inversion symmetry breaking in bulk strained semiconductors.
\end{abstract}


\maketitle

The polarization and coherent manipulation of electron spins are important steps towards the realization of spin-based information processing~\cite{wolf01}. While electron spins can be manipulated by magnetic fields, electrical control is desirable as it offers the potential for high-speed manipulation and local gates. Spin-orbit interactions in semiconductors offer momentum-dependent effective magnetic fields that can be used for electrical control~\cite{datta,physics}.

Momentum {\bf k}-dependent spin splittings arise from the spin-orbit interaction and the breaking of spatial inversion symmetry. The lack of an inversion center in zincblende crystal structures results in a bulk inversion asymmetry (BIA) and the Dresselhaus field~\cite{dresselhaus}. Structural inversion asymmetry (SIA) along the growth direction in heterostructures results in the Rashba field~\cite{rashba}. These internal effective magnetic fields have different {\bf k}-dependence and can be distinguished by changing the direction of the carrier drift momentum. Measurements as a function of electric field direction have mapped the {\bf k}-linear Rashba and Dresselhaus fields in quantum wells~\cite{meier}.

Strain breaks spatial inversion symmetry, which results in additional {\bf k}-dependent spin splittings~\cite{opticalorientation,larocca,bernevig}. Strain-induced spin precession and characterization of {\bf k}-linear spin splitting have been conducted in lattice-mismatched heterostructures~\cite{kato04} and using a mechanical vise~\cite{crooker05,beck,sih06}. Measurements on lattice-mismatched heterostructures observed both BIA- and SIA-type splitting, but the splitting did not exhibit a clear trend as a function of measured strain~\cite{kato04}. A subsequent theoretical analysis proposed that variations in strain relaxation during growth could result in different {\bf k}-linear BIA-type splitting~\cite{bernevig}. Measurements conducted using a mechanical vise found that uniaxial strain along $\langle$110$\rangle$ introduces a SIA-type splitting~\cite{crooker05,beck}.

In this paper we investigate the momentum dependence of strain-induced spin-orbit splittings in In$_{0.04}$Ga$_{0.96}$As epilayers. The InGaAs epilayers are grown on GaAs substrates, which introduces biaxial compressive strain, and the sample structure was designed to reduce inhomogeneous effects related to strain relaxation. Measurements of internal effective magnetic fields as a function of electron spin drift velocity along [100], [010], [110] and [1$\overline{1}$0] enable separation of BIA and SIA-type spin splittings that arise from biaxial and uniaxial strain.

For {\bf k} in the plane perpendicular to the growth direction [001], the spin-splitting Hamiltonians take the form:
\begin{eqnarray}
\mbox{H}_{D} = \lambda\left(\sigma_{x}k_{x}k_{y}^{2}-\sigma_{y}k_{y}k_{x}^{2}\right)
\end{eqnarray}
\begin{eqnarray}
\mbox{H}_{R} = \alpha\left(k_{y}\sigma_{x} - k_{x}\sigma_{y}\right)
\end{eqnarray}
\begin{eqnarray}
\mbox{H}_{1} = D\left(\epsilon_{zz} - \epsilon_{xx}\right)\left(\sigma_{x}k_{x} - \sigma_{y}k_{y}\right)
\end{eqnarray}
\begin{eqnarray}
\mbox{H}_{2} = \frac{C_{3}\epsilon_{xy}}{2}\left(k_{y}\sigma_{x} - k_{x}\sigma_{y}\right)
\end{eqnarray}

H$_{D}$ and H$_{R}$ are the Dresselhaus and Rashba Hamiltonians, respectively, while H$_{1}$ and H$_{2}$ are two additional {\bf k}-linear terms due to strain~\cite{opticalorientation,larocca,bernevig}. Here $x$, $y$, and $z$ denote the [100], [010], and [001] crystal axes, $\sigma_{i}$ denotes the $i^{th}$ Pauli matrix, $\epsilon_{ij}$ are the components of the strain tensor with $\epsilon_{xy} = \epsilon_{yx}$ and $\epsilon_{xx} = \epsilon_{yy}$, and $\lambda$, $\alpha$, $D$, and $C_{3}$ are material constants. H$_{R}$ and H$_{2}$ have the same direction dependence on momentum, while H$_{1}$ has the same form as the linear Dresselhaus field for a two-dimensional system with quantum confinement along [001]~\cite{winkler}. H$_{2}$ accounts for the spin splitting introduced by uniaxial strain along $\langle$110$\rangle$~\cite{crooker05,beck,hruska}, and H$_{1}$ has been proposed~\cite{bernevig} to explain the BIA-type splitting in Ref.~[\onlinecite{kato04}], despite earlier work~\cite{opticalorientation} that argued that H$_{1}$ should be small. The directions of the SIA-type (H$_{R}$ and H$_{2}$) and BIA-type (H$_{D}$ and H$_{1}$) fields are shown in Fig.~1(b). For {\bf k} along [110] and [1$\overline{1}$0], the SIA and BIA fields are both perpendicular to {\bf k} and parallel to each other, while for {\bf k} along [100] and [010], the BIA fields are parallel to {\bf k} while the SIA fields are perpendicular to {\bf k}.

In general, H$_{D}$ is cubic in {\bf k} while H$_{R}$, H$_{1}$, and H$_{2}$ are linear in {\bf k}. Thus the spin splitting can be described by combinations of {\bf k}-linear and {\bf k}-cubic terms of the form $\Delta = \alpha v_{d}^{3} + \beta v_{d}$, where $v_{d}$ is the in-plane drift velocity defined by $\hbar k_{d} = m^{\ast}v_{d}$ with $m^{\ast}$ the conduction band electron effective mass and $\hbar k_{d}$ the drift momentum. Yet along the [010] axis, $k_{x}$ = 0 and $k_{y}$ = $|${\bf k}$_{d}|$, so H$_{D}$ = 0 for bulk systems. Similarly, H$_{D}$ = 0 for {\bf k} along [100]. On the other hand, for momentum along [110], $k_{x} = k_{y} = \frac{k_{d}}{\sqrt{2}}$, and H$_{D}$ reduces to H$_{D}$ $=$ $\frac{\lambda k_{d}^{3}}{2\sqrt{2}}\left( \sigma_{x} - \sigma_{y} \right)$.

\begin{figure}
\includegraphics[width=0.47\textwidth]{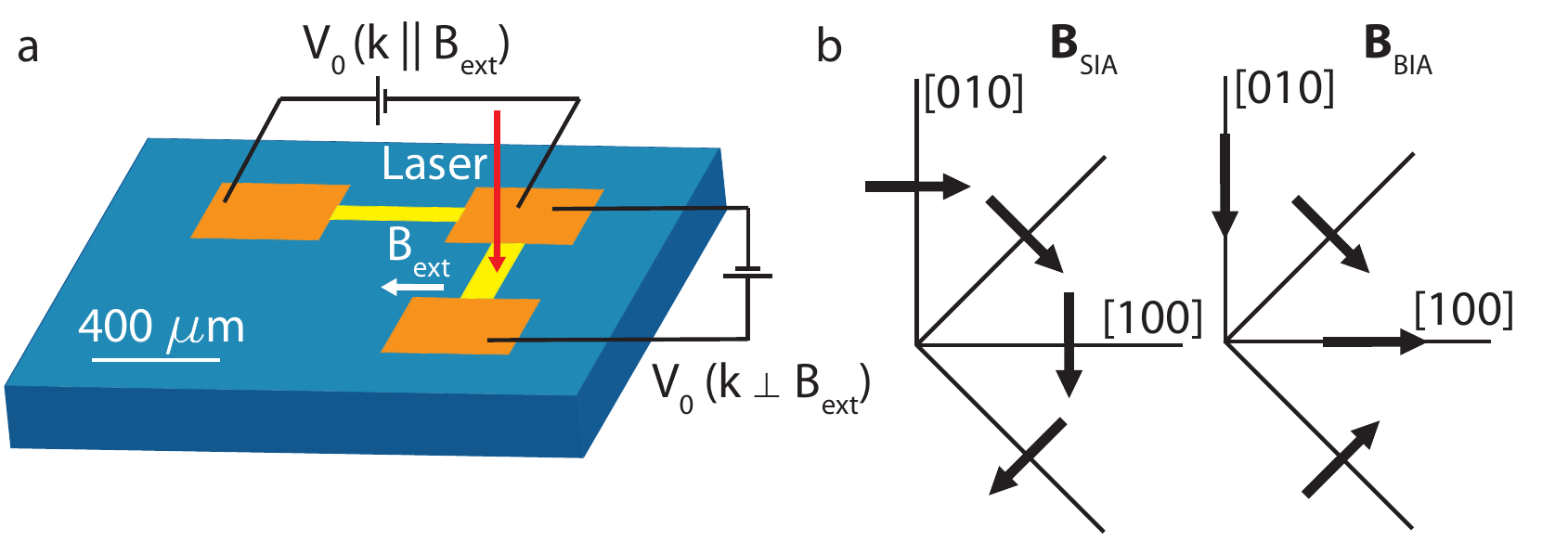}
\caption{\label{fig1} (Color) (a) Experimental setup. InGaAs channels and contacts are shown in yellow and orange respectively. Voltage is applied such that the net electron drift momentum is either perpendicular to or parallel to the external magnetic field {\bf B}$_{\mathrm{ext}}$. The red arrow denotes the laser propagation direction. (b) Directions of momentum-dependent effective internal fields due to spin-orbit interactions from structural inversion asymmetry (left) and bulk inversion asymmetry (right). }
\end{figure}

The momentum-dependent effective magnetic fields described above were measured using pump-probe optical techniques. Optical orientation of electron spins results in an out-of-plane spin polarization, which can precess about in-plane internal and applied external magnetic fields. Measurements were performed on Si-doped In$_{0.04}$Ga$_{0.96}$As epilayers (doping concentration $n$ = $3 \times 10^{16}$ cm$^{-3}$). The thickness of the InGaAs layer is 500 nm, grown above a 300 nm growth interrupted GaAs buffer layer on a (001) GaAs substrate and capped with 100 nm undoped GaAs. X-ray diffraction measurements show that the InGaAs epilayer is lattice-matched to the GaAs substrate and exhibits minimal strain relaxation. Samples were fabricated, consisting of two perpendicular InGaAs channels [Fig.~1(a)]. Channels were aligned along [110], [1$\overline{1}$0], [100] and [010]. The channels were each 400 $\mu$m long and 100 $\mu$m wide with 400 x 400 $\mu$m$^{2}$ ohmic contacts at each end. Wires were soldered to each of the contacts and connected to an external voltage source. The samples were mounted in a liquid helium flow cryostat with measurements made at temperature $T$ = 30 K. Care was taken during sample mounting to minimize introducing additional strain.

Measurements were conducted using a mode-locked Ti:Sapphire laser with a repetition rate of 76 MHz tuned to the band edge of the InGaAs epilayer ($\lambda$ = 848 nm). The laser was separated into a pump and probe pulse with a temporal separation controlled by a mechanical delay line. The pump and probe beams were modulated by a photoelastic modulator (PEM) and optical chopper respectively for cascaded lock-in detection. An external magnetic field was applied in the plane of the layer and varied from -40 to 40 mT. The inclusion of a motor driven steering mirror in the pump path allowed control of the pump-probe spatial separation on the sample. Field scans were taken at 5 $\mu$m intervals over a 40 $\mu$m range (roughly $\pm$20 $\mu$m from the center of the electron spin packet). By applying an external voltage to the contacts, the electron drift momentum was varied along each channel. Measurements were performed with the channels oriented both parallel with and perpendicular to the applied magnetic field to measure both the magnitude and direction of the internal magnetic fields.

Faraday rotation (FR) of the probe pulse can be described by:
\begin{eqnarray}
\theta_{F} = A e^{-\frac{\Delta\text{t}}{T_{2}^{\ast}}}
\cos\left(\frac{\text{g}\mu_{B}\Delta\text{t}}{\hbar} \vert {\bf B}_{\mathrm{ext}} + {\bf B}_{\mathrm{int}} \vert \right)
\end{eqnarray}
where $A$ is the FR amplitude, g is the electron g factor, $\mu_{B}$ is the Bohr magneton, $T_{2}^{\ast}$ is the inhomogeneous dephasing time, $\Delta$t is the pump-probe time delay, and {\bf B}$_{\mathrm{ext}}$ is the applied external magnetic field. Time-resolved FR measurements~\cite{crooker} find that g = 0.51 and $T_{2}^{\ast}$ = 7.8 ns. The direction and magnitude of the internal magnetic field {\bf B}$_{\mathrm{int}}$, which is proportional to the spin splitting $\Delta$ = $\text{g}\mu_{B} \vert {\bf B}_{\mathrm{int}} \vert$, is determined from fits of the FR signal to Eq. 5 with errors less than 1$\%$. As shown in Fig.~2(a), the component of {\bf B}$_{\mathrm{int}}$ that is parallel to {\bf B}$_{\mathrm{ext}}$ causes an overall shift in the field-dependent signal, and the component that is perpendicular to {\bf B}$_{\mathrm{ext}}$ changes the magnitude of the center peak. B$_{\parallel}$ (B$_{\perp}$) is defined as the component of {\bf B}$_{\mathrm{int}}$ that is parallel (perpendicular) to ${\bf k}_{d}$, as determined from field scans based on the channel orientation. Measurements were taken with {\bf k} $\perp$ {\bf B}$_{\mathrm{ext}}$ and {\bf k} $\parallel$ {\bf B}$_{\mathrm{ext}}$ to determine both the sign and magnitude of the spin splitting. The magnitude of the drift momentum was varied by changing the potential applied to the channels, in the range between -2.0V and 2.0V. Scans with varying pump-probe spatial separations were taken to characterize the internal field and drift velocity of the optically injected spin packet [Fig. 2]. The drift velocity $v_{d}$ of the spin packet at each voltage is the measured spatial drift of the spin packet center $x_{c}$ during the pump-probe time delay $\Delta t = 13$ ns.

\begin{figure}
\includegraphics[width=0.47\textwidth]{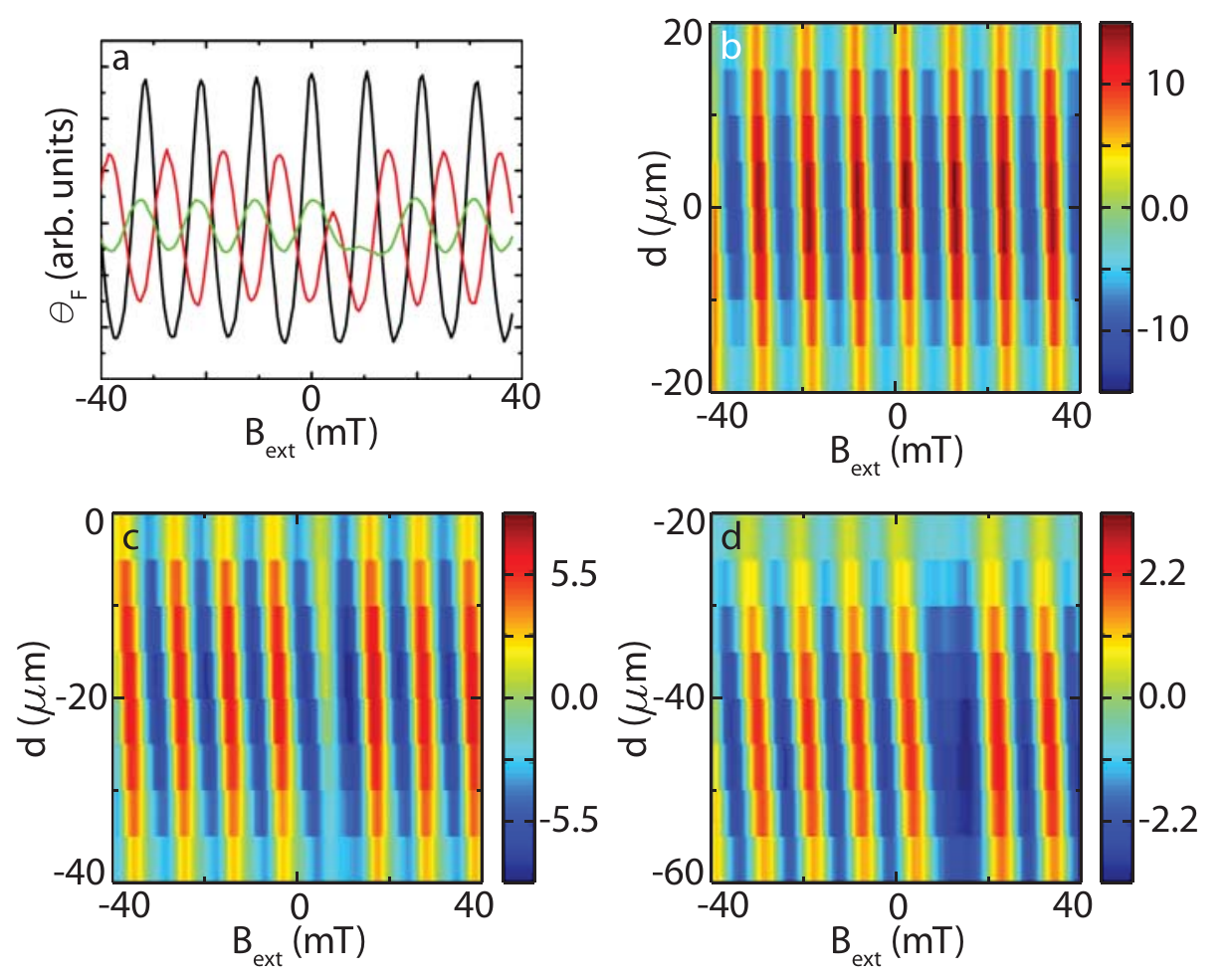}
\caption{\label{fig2} (Color) (a) Faraday rotation $\theta_{F}$ as a function of external magnetic field {\bf B}$_{\mathrm{ext}}$ for 0.0V (black), 1.0V (red) and 2.0V (green) measured at the center of the spin packet. The [010] channel was oriented perpendicular to {\bf B}$_{\mathrm{ext}}$, so the shift and decrease in the center peak give the values of B$_{\perp}$ and B$_{\parallel}$ respectively. (b)-(d) Faraday rotation as a function of {\bf B}$_{\mathrm{ext}}$ and pump-probe separation $d$ for (b) 0.0V, (c) 1.0V, and (d) 2.0V. The data was measured at $T$ = 30 K with a pump-probe time delay of $\Delta$t = 13 ns. }
\end{figure}

For the [010] channel oriented perpendicular to the external magnetic field, values of B$_{\perp}$ and B$_{\parallel}$ are displayed for various contact voltages as a function of pump-probe spatial separation in Fig. 3. The amplitude of the FR is also fit for each scan to determine the position of the spin packet center $x_{c}$ at each voltage [Fig.~3(c)]. The spatial dependence of {\bf B}$_{\mathrm{int}}$ is due to both drift and diffusion of the spin packet~\cite{kato04}. For the [010] channel, it is expected that {\bf B}$_{\mathrm{int}}$ has a linear dependence on {\bf k} because H$_{D}$ = 0. For each voltage, a linear fit is used to determine the value of {\bf B}$_{\mathrm{int}}$ at $x_{c}$. These values are fit to linear functions of $v_{d}$ [Fig. 3(d)] to determine the values of $\beta_{\perp}$ and $\beta_{\parallel}$.

\begin{table*}
\renewcommand{\arraystretch}{1.5}
\begin{ruledtabular}
\begin{tabular}{|c|c|c|c|c|c|c|c|c|}
  \hline
    & \multicolumn{4}{c|}{$k\parallel B_{ext}$} & \multicolumn{4}{c|}{$k\perp B_{ext}$} \\
  \hline
  Channel & $\beta_{\parallel}$ & $\beta_{\perp}$ & $C_{3}\epsilon_{xy}$/$\hbar$ & $D(\epsilon_{zz}-\epsilon_{xx})$/$\hbar$ & $\beta_{\parallel}$ & $\beta_{\perp}$ & $C_{3}\epsilon_{xy}$/$\hbar$ & $D(\epsilon_{zz}-\epsilon_{xx})$/$\hbar$ \\
  \hline
  [100] & 35 & -84 & -230 & 47 & 35 & -110 & -300 & 47 \\
  \hline
  [010] & -35 & -100 & -280 & 48 & -41 & -91 & -250 & 56 \\
  \hline
  [110] & 0 & -14 & -180 & 70 & 0 & -20 & -150 & 50 \\
  \hline
  [1$\overline{1}$0] & 0 & -120 & --- & --- & 0 & -94 & --- & --- \\
  \hline
\end{tabular}
\end{ruledtabular}
\caption{\label{tab:tc} Measured $\beta_{\parallel}$ and $\beta_{\perp}$ for each channel and orientation expressed in units of neV ns $\mu$m$^{-1}$. $\beta_{\parallel}$ represents the BIA spin splitting, while $\beta_{\perp}$ represents either the SIA ([100] and [010] channels), SIA plus BIA ([110]) or SIA minus BIA ([1$\overline{1}$0]) spin splitting. Calculated strain parameters C$_{3}$$\epsilon_{xy}$/$\hbar$ and D($\epsilon_{zz}-\epsilon_{xx}$)/$\hbar$ are in units of m/s. }
\end{table*}

\begin{figure}
\includegraphics[width=0.47\textwidth]{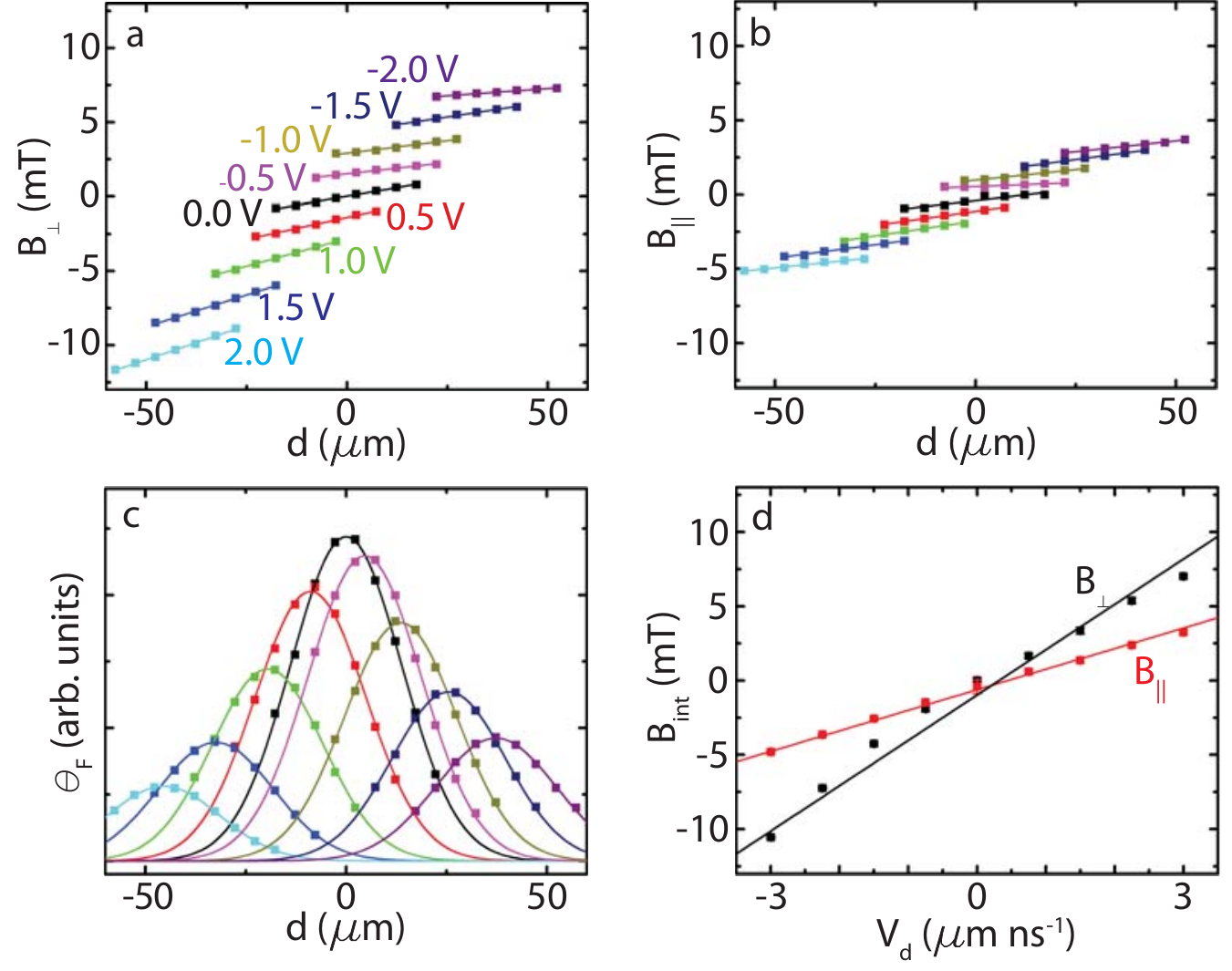}
\caption{\label{fig3} (Color) Internal magnetic fields (a) perpendicular to and (b) parallel to the channel direction for the [010] channel at $T$ = 30 K with a pump-probe time delay of $\Delta$t = 13 ns. For these measurements, the channel was oriented perpendicular to the applied magnetic field ({\bf k} $\perp$ {\bf B}$_{\mathrm{ext}}$). Applied voltages are the same for (a)-(c). Lines in (a) and (b) are linear fits. (c) Faraday rotation amplitude as a function of pump-probe separation for various potential differences used to determine the spin packet center $x_{c}$. (d) B$_{\perp}$ (black) and B$_{\parallel}$ (red) evaluated at $x_{c}$ as a function of spin packet drift velocity. The slope of the linear fit gives the value of $\beta/g\mu_{B}$. Errors in (a)-(c) are smaller than the data points. }
\end{figure}

Table 1 shows a summary of the measured values of $\beta_{\perp}$ and $\beta_{\parallel}$ for all four channel directions and both measurement orientations, {\bf k} $\parallel$ {\bf B}$_{\mathrm{ext}}$ and {\bf k} $\perp$ {\bf B}$_{\mathrm{ext}}$. All of the measurements are well described by a linear dependence of {\bf B}$_{\mathrm{int}}$ with {\bf k}, so $\beta$ = $\Delta$/$v_{d}$. Measurements with {\bf k} $\parallel$ {\bf B}$_{\mathrm{ext}}$ and {\bf k} $\perp$ {\bf B}$_{\mathrm{ext}}$ show reasonable agreement, but with noticeable discrepancies that are larger than the measurement error, which is typically on the order of 1 $\--$ 5 $\%$. Additional strain introduced during sample mounting and cool-down could possibly account for these discrepancies.  For the [100] and [010] channels for which the SIA- and BIA-type spin splittings can be separated, we found that the SIA splitting ($\beta_{\perp}$) is consistently 2 or 3 times larger than the BIA splitting ($\beta_{\parallel}$). For the [010] channel, the value of $\beta_{\perp}$ can be attributed to H$_{R}$ and H$_{2}$ and $\beta_{\parallel}$ to H$_{1}$, since H$_{D} = 0$ for {\bf k} along [010]. Also, as expected from the form of H$_{1}$ [Eq. 3], the BIA-type splitting ($\beta_{\parallel}$) has similar magnitude but opposite sign for [100] and [010], which justifies the assumption that $\epsilon_{xx}$ $=$ $\epsilon_{yy}$ for this sample.

Although the measurements cannot distinguish between the contributions of H$_{R}$ and H$_{2}$, previous measurements indicate that the effect of H$_{2}$ is greater than that of H$_{R}$ for similar bulk epilayers.  Measurements~\cite{kato04} on unstrained n-doped GaAs epilayers found $\beta < 10 \mbox{ neV ns } \mu \mbox{m}^{-1}$, and that the measured spin splitting upon applied strain could be characterized using H$_{2}$ and without requiring H$_{R}$~\cite{crooker05,beck,sih06}.

From Eqs.~3 and 4 and the measured values of $\beta$, we calculate the sample parameters C$_{3}$$\epsilon_{xy}$/$\hbar$ and D($\epsilon_{zz}-\epsilon_{xx}$)/$\hbar$ assuming no contribution from H$_{D}$ and H$_{R}$ and $m^{\ast} = 0.065 m_{0}$~\cite{vurgaftman}, where m$_{0}$ is the free electron mass.  From the measurements on the [100] and [010] channels, C$_{3}$$\epsilon_{xy}$/$\hbar$ has an average value of -270 m/s, and D($\epsilon_{zz}-\epsilon_{xx}$)/$\hbar$ = 50 m/s. From Ref.~[\onlinecite{bernevig}], C$_{3}$/$\hbar$ = $8 \times 10^{5}$ m/s and D/$\hbar$ = (0.5 - 1.5) $\times 10^{4}$ m/s. From the literature value of C$_{3}$/$\hbar$ and the range of values of C$_{3}\epsilon_{xy}$/$\hbar$ measured here, $\epsilon_{xy}$ = -0.029 $\--$ -0.038$\%$. Therefore, the variation in our measurements indicates that the strain changes by an amount less than $\vert \Delta \epsilon_{xy} \vert = 1$ $\times$ $10^{-4}$ between different sample mountings and cool-downs. For pseudomorphic In$_{0.04}$Ga$_{0.96}$As on GaAs~\cite{jain}, $\epsilon_{xx}$ = -0.29$\%$ and $\epsilon_{zz}$ = 0.26$\%$, which yields D/$\hbar$ = $0.9 \times 10^{4}$ m/s, which is in the same range as previously reported values~\cite{kato04,bernevig}. The data and analysis show that the measured spin splittings are consistent with H$_{1}$ and H$_{2}$ and that H$_{2}$ can have comparable magnitude to H$_{1}$, which contradicts the argument that H$_{1}$ is small and can be neglected~\cite{opticalorientation,hruska}.

\begin{figure}
\includegraphics[width=0.47\textwidth]{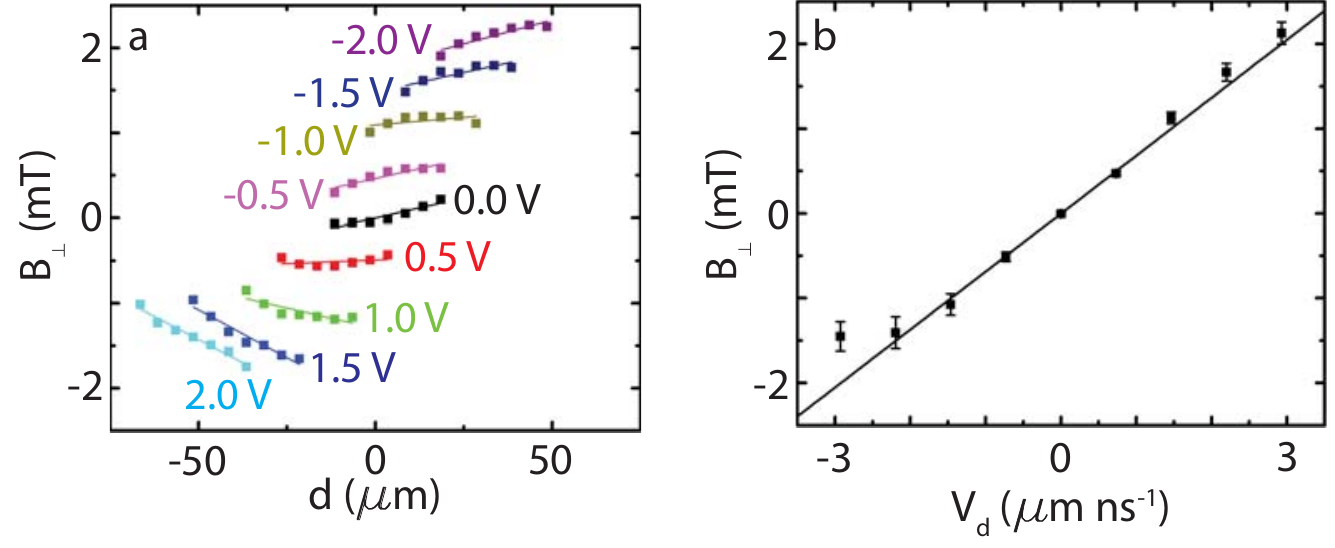}
\caption{\label{fig4} (Color) (a) Internal magnetic field perpendicular to {\bf k} measured with {\bf k} $\perp$ {\bf B}$_{\mathrm{ext}}$ for the [110] channel at $T$ = 30 K as a function of pump-probe separation $d$ and with a pump-probe time delay of $\Delta$t = 13 ns. (b) Internal magnetic field evaluated at the spin packet center as a function of electron spin packet velocity. The slope of the linear fit gives the value of $\beta$/g$\mu_{B}$. Errors in (a) are smaller than the data points.  }
\end{figure}

Unexpectedly, the values of $\beta_{\perp}$ for the [1$\overline{1}$0] and [110] channels do not match the difference and sum, respectively, of the SIA and BIA terms from the other channels. Therefore, measurements along [110] and [1$\overline{1}$0] may not be sufficient to accurately characterize H$_{1}$ and H$_{2}$. For [110], $\beta_{\perp}$ = -14 neV ns $\mu$m$^{-1}$ for {\bf k} $\parallel$ {\bf B}$_{\mathrm{ext}}$ and -20 neV ns $\mu$m$^{-1}$ for {\bf k} $\perp$ {\bf B}$_{\mathrm{ext}}$. These values are both smaller in magnitude than the average sum $\left(\beta_{\perp} + \vert \beta_{\parallel} \vert \right)$ = -58 neV ns $\mu$m$^{-1}$ from measurements on the [100] and [010] channels. Similarly, for [1$\overline{1}$0], $\beta_{\perp}$ = -120 neV ns $\mu$m$^{-1}$ for {\bf k} $\parallel$ {\bf B}$_{\mathrm{ext}}$ and -94 neV ns $\mu$m$^{-1}$ for {\bf k} $\perp$ {\bf B}$_{\mathrm{ext}}$, and both values are smaller in magnitude than the average difference $\left(\beta_{\perp} - \vert \beta_{\parallel} \vert \right)$ = -130 neV ns $\mu$m$^{-1}$. We also calculate C$_{3}$$\epsilon_{xy}$/$\hbar$ and D($\epsilon_{zz}-\epsilon_{xx}$)/$\hbar$ assuming that H$_{1}$ $\propto$ $(\beta_{[110]}-\beta_{[1\overline{1}0]})/2$ and H$_{2}$ $\propto$ $(\beta_{[110]}+\beta_{[1\overline{1}0]})/2$ and obtain a smaller value for C$_{3}$$\epsilon_{xy}$/$\hbar$. The calculated C$_{3}$$\epsilon_{xy}$/$\hbar$ and D($\epsilon_{zz}-\epsilon_{xx}$)/$\hbar$ values are included in Table 1. We hypothesize that higher order terms, including the additional {\bf k}-cubic Dresselhaus term (H$_{D}$) for [110] and [1$\overline{1}$0], could account for this discrepancy.

We also observe unexpected behavior in the measured slope of {\bf B}$_{\mathrm{int}}$ as a function of pump-probe spatial separation and voltage for the [110] channel. The slope of {\bf B}$_{\mathrm{int}}$ at each voltage is due to diffusion and the distribution of drift velocity of the spin packet from the finite spot size (~30 x 10 $\mu$m) of the pump and probe beams~\cite{kato04}. Electrons at the leading edge of the packet have a higher average velocity than those in the back and thus we expect a non-zero and positive slope. However, as shown in Fig.~4(a), for scans in the voltage range 1.0 - 2.0V for the [110] sample, surprisingly, a negative slope is observed. This was found from repeated measurements on different spots along the channel and is therefore not thought to be a result of any interface reflection effects. The measurements of $\beta$ do not seem to be affected, as the value for {\bf B}$_{\mathrm{int}}$ for the center of the spin packet at each voltage still has a linear dependence on spin drift velocity, as shown in Fig.~4(b). Measurements for both configurations {\bf k} $\parallel$ {\bf B}$_{\mathrm{ext}}$ and {\bf k} $\perp$ {\bf B}$_{\mathrm{ext}}$ show reasonable agreement, with values for $\beta_{\perp}$ of -14 and -20 neV ns $\mu$m$^{-1}$, respectively. The smaller spin splitting could possibly explain why this behavior is only apparent in the [110] sample. Larger values of $\beta$ could overwhelm this effect in other channels.

This material is based upon work supported by the National Science Foundation under Grants No. ECCS-0844908 and DMR-0801388 and the Horace H. Rackham School of Graduate Studies. Sample fabrication was performed in the Lurie Nanofabrication Facility, part of the NSF funded NNIN network.

\end{document}